*Does nitrate deposition following astrophysical ionizing radiation events pose an additional threat to amphibians?*

Brian C. Thomas[1], Michelle D. Honeyman[2]

[1]Department of Physics and Astronomy, Washburn University, Topeka, Kansas
1700 SW College Ave., Topeka, Kansas, 66621, USA; 785-670-2144;
brian.thomas@washburn.edu (to whom correspondence should be addressed)

[2]Department of Biology, Washburn University, Topeka, Kansas

As diversity in amphibian species declines, the search for causes has intensified. Work in this area has shown that amphibians are especially susceptible to the combination of heightened UVB radiation and increased nitrate concentrations. Various astrophysical events have been suggested as sources of ionizing radiation that could pose a threat to life on Earth, through destruction of the ozone layer and subsequent increase in UVB, followed by deposition of nitrate. In this study, we investigate whether the nitrate deposition following an ionizing event is sufficiently large to cause an additional stress beyond that of the heightened UVB previously considered. We have converted predicted nitrate depositions to concentration values, utilizing data from the New York State Department of Environmental Conservation Acid Rain Monitoring Network web site. Our results show that the increase in nitrate concentration in bodies of water following the most intense ionization event likely in the last billion years would not be sufficient to cause a serious additional stress on amphibian populations and may actually provide some benefit by acting as fertilizer.

1. Introduction

In previous work (Melott *et al.*, 2005; Thomas *et al.*, 2005, 2007, 2008; Thomas and Melott, 2006; Ejzak *et al.*, 2007) we have evaluated the biological impact of astrophysical ionizing radiation events, including gamma-ray bursts, supernovae and solar flares. We have computed relative DNA damage caused by heightened solar UVB (290-315 nm) irradiation after depletion of stratospheric ozone. We have also computed deposition rates of nitrate (the result of removal of nitrogen oxides from the atmosphere as $HNO_3$) after particularly energetic events. In this study, we investigate the possibility that this nitrate deposition could result in an increased biological impact, in combination with already studied UVB irradiation. The primary question is whether ionizing radiation events such as those studied in the past would produce enough of an increase in nitrate concentrations to significantly enhance the damage caused by UVB alone. That is, is it possible that events that did not appear to be significantly harmful could actually be more dangerous when the impact of both UVB and nitrate concentrations are considered as a combined stressor?

The combination of increased UVB exposure at the Earth's surface and enhanced nitrate deposition as the atmosphere recovers is a generic result of astrophysical ionizing radiation events. It is important to note that a single event would cause both effects; ionization leads to formation of nitrogen oxides that catalytically deplete ozone and then are removed from the atmosphere (over several months' time) as $HNO_3$, leading to deposition of $NO_3^-$ (Thomas *et al.*, 2005). Any event that yields significant ionization of the atmosphere will lead to both effects.

There is considerable evidence that anthropogenic ozone depletion and resulting increased UVB is contributing to declines in amphibians and other biota (Blaustein *et al.,* 1994; Blaustein and Wake, 1995; Belden and Blaustein, 2002; Bancroft *et al.,* 2007). These studies show that UVB alone is a significant stress on amphibian and other populations. In addition, studies have shown that the combination of stressors such UVB irradiation, increased nitrate concentrations and decreased pH has a greater effect than the individual stressors, or even a simple sum of the effects of each separately (Hatch and Blaustein, 2000, 2003; Boone *et al.*, 2007). Amphibians are particularly sensitive to increased levels of both UVB and nitrate because they lay their eggs in shallow waters. Hatch and Blaustein (2000, 2003) report reduced survival, mass, length and activity of larvae of Cascades frogs (*Rana cascadae*), Pacific treefrogs (*Hyla regilla*), and long-toed salamanders (*Ambystoma macrodactylum*), due to the combination of increased UVB and nitrate concentrations. Johansson *et al*. (2001) found that nitrate alone is not a significant threat to the development of larvae of the common frog (*Rana temporaria*). Therefore, in this study we consider nitrate deposition as an additional stressor to amphibian populations already impacted by heightened UVB levels after an astrophysical ionizing radiation event.

2. Methods

Previous studies have reported computed deposition rates of nitrate following astrophysical ionizing radiation events, in particular gamma-ray bursts (GRBs). These events were modeled using the NASA Goddard Space Flight Center 2D atmospheric chemistry and dynamics model. The model and results of simulation runs with nitrate deposition rates for various events are described in Thomas *et al*.

(2005) and Melott *et al.* (2005). Our use of this model for other types of events is described in (Ejzak *et al.*, 2007; Thomas *et al.*, 2007, 2008)

For this study, we are interested in evaluating the maximum possible impact of nitrate as an additional stressor, specifically on amphibian populations, which are known to be particularly sensitive to the combination of increased UVB and nitrate concentrations. To this end, we use the largest nitrate deposition rate from a suite of simulation runs reported in Thomas *et al.* (2005), for a GRB delivering a fluence of 100 kJ m$^{-2}$. This maximum value is 3 x 10$^{-9}$ g m$^{-2}$ s$^{-1}$, for a burst occurring over the North Pole in September (the Autumnal Equinox). Maximum deposition occurs over a few months, between 18 and 24 months after the burst, at latitudes between 30 and 50 degrees North.

In order to compare to amphibian studies, we must convert this deposition rate to a concentration value in water. Such conversion is subject to a variety of uncertainties including where the deposition occurs, what types of bodies of water are considered, how the nitrate is transported from the deposition site to the water (or if it is directly deposited), reactions with various compounds in the deposition region (either on land or in water), etc. There is no way for us to accurately model such complex and variable transport and conversion in a theoretical way.

We may make an initial estimate by multiplying the deposition rate by $3.15 \times 10^7$ s to convert to a yearly value, yielding 0.09 g m$^{-2}$. Multiplying by the area of deposition yields $6.12 \times 10^{12}$ g, which is about $10^{11}$ moles of $NO_3^-$. Converting to a concentration value requires assuming a water volume. The available surface water in the region of deposition may be around $10^{16}$ L, which gives a concentration of around 0.6 mg L$^{-1}$. One can also consider pH values, which should be around 5, and hence not significant.

To obtain a more accurate estimate of concentration, we chose to develop an empirical conversion factor using field measurements of both nitrate deposition and concentration from existing acid rain studies.

We utilized data gathered by the New York State Department of Environmental Conservation Acid Rain Monitoring Network, retrieved from their website (http://www.dec.ny.gov/chemical/283.html). The network has been in operation since 1987 and consists of 21 monitoring sites in both rural and urban areas. Several different analyses are made of samples collected at each site. Most important for our purposes are measurements of nitrate ion ($NO_3^-$) deposition (given in kg ha$^{-1}$) and concentration (given in mg L$^{-1}$). Comparison of these two measured values allows us to determine an empirical conversion factor between deposition and concentration. This conversion can then be applied to the deposition value given by our modeling of a GRB or other event. Once we have a concentration value for an event, we can compare that concentration, in combination with UVB irradiance also computed for the event, to the same quantities reported in studies of amphibians. Thus we can determine whether the nitrate deposition following an astrophysical radiation event is a significant additional stressor for these populations or not.

It is important to note that this procedure yields only an upper limit on the impact. Samples from monitoring stations are collected in containers which are exposed directly to rainfall or which collect runoff from an artificial surface. These samples are not filtered through soil, diluted in rivers or ponds, etc., as would normally be the case for deposition that affects amphibians in natural environments. Therefore, the concentration values derived from these measurements are likely to be larger, for a given deposition rate, than they would be in nature. While this certainly represents a large uncertainty, the nature of our results is such that an upper limit is appropriate and therefore this method is sufficient for the present study.

While pH has been shown to be an additional stressor on amphibian populations, we have not considered that factor in this work. Data on pH is available from the Acid Rain Monitoring Network, but nitrate is not the only contributor; sulfate is a much more important source of acidity in the measurements. Therefore, since it is impossible to separate out the sole contribution of nitrate, we have chosen to include only concentration as an additional stress factor, and not pH.

We have compiled yearly values for nitrate deposition and concentration from 18 monitoring sites, over 10 years (between 1994 and 2004). For each data point (site and year) we compute a conversion factor equal to the concentration (in mg L$^{-1}$) divided by the deposition (in kg ha$^{-1}$). The final conversion factor value will then multiply our modeled deposition value to yield concentration. The conversion factors for each year at a given site are then averaged, and then averaged again across all sites. An estimate of the uncertainty in the final value is given by computing the standard deviations for the first set of averages, and then computing the root of the sum of the squares of these to get a final single error estimate for the final conversion factor value. See Table 1. It should be emphasized that this is simply an estimate of the statistical uncertainty, and in no way includes the much larger systematic uncertainties which apply in comparing these values to what would be the case in a natural setting.

3. **Results**

Following the procedure described above, we arrive at a final conversion factor value of $0.10 \pm 0.08$ mg ha $L^{-1}$ $kg^{-1}$. Our modeled deposition rate ($3 \times 10^{-9}$ g $m^{-2}$ $s^{-1}$) is multiplied by $3.15 \times 10^{7}$ s to convert to a total yearly value, yielding 0.09 g $m^{-2}$ = 0.9 kg $ha^{-1}$. Finally, multiplying by our conversion factor gives an expected concentration of 0.09 mg $L^{-1}$. If we take the high end of conversion factor range given by the statistical uncertainty, we arrive at a concentration of 0.16 mg $L^{-1}$.

This concentration can be compared to values in studies linking UVB and nitrate enhancement to amphibian population decline. Hatch and Blaustein (2000, 2003) use nitrate concentrations between 5 and 20 mg $L^{-1}$. Our modeled concentrations are roughly two orders of magnitude below these values.

4. **Discussion**

Given that uncertainties in converting from deposition to concentration favor lower concentration values in a natural setting, we may safely say that the example ionization event considered here will not significantly impact amphibian populations through increased nitrate concentrations. This example event is at the high end of the likely damage range, when compared to other possible sources of ionizing radiation (e.g. supernovae, solar flares, etc.) since other events deliver comparable or lower ionizing fluence.

Hatch and Blaustein (2003) found that UVB alone did not have an impact on the organisms studied, while the combination of nitrate and UVB did have an impact on survivability. It is important to note that in those experiments the UVB irradiance ranged up to a maximum of about 20 $\mu W\ cm^{-2}$, while in our previous studies of GRBs, we computed surface irradiance values up to about 500 $\mu W\ cm^{-2}$. With this level of irradiance, and given studies showing that UVB alone does adversely affect some amphibian species (e.g. Blaustein and Wake, 1994), we would still expect to see an impact on this group of organisms, even without a significant nitrate enhancement. Of course, it is known that UVB broadly impacts a variety of organisms (Bancroft *et al.*, 2007) and our previous work has also included computations of DNA damage (using the weighting function of Setlow, 1974) that show a significant impact on organisms which are not shielded.

It is interesting to consider what magnitude event would be necessary in order to produce nitrate concentrations on the order of that used in the Hatch and Blaustein (2000, 2003) studies. Given that odd nitrogen production in the atmosphere by gamma-rays scales linearly with the received ionizing fluence (Thomas *et al.*, 2005, Ejzak *et al.*, 2007), and assuming that nitrate deposition scales roughly linearly with odd nitrogen production, we may estimate that a fluence of some 10 MJ m$^{-2}$ would be required to produce nitrate concentrations of order 10 mg L$^{-1}$. An event delivering 100 kJ m$^{-2}$ (as was assumed above) is likely to occur roughly every billion years, at a distance of about 2 kpc, assuming a "typical" burst of power 5 x 10$^{44}$ W and duration 10 s, as described in Thomas *et al.* (2005). A similar event delivering 100 times that fluence would correspond to a distance of about 200 pc. This is about one-third the distance to the nearest probable burst in the last billion years based on the least conservative assumptions. Of course, an event of this magnitude would decimate the stratospheric ozone layer and the resulting increase in UVB would have a dramatic impact, even in the absence of any enhancement in nitrate concentrations.

It has been noted previously that increased nitrate deposition after an astrophysical ionizing radiation event may actually benefit primary producers by acting as fertilizer (Melott *et al.*, 2005; Thomas *et al.*, 2005). Small amounts of nitrate may thereby indirectly *benefit* amphibian populations by increasing productivity of food sources. Increased levels of nitrate in aquatic environments provides nutrients for the growth of algae, etc. (Shi *et al.*, 2005; Smith *et al.*, 1999), which in turn provides food for larval amphibians and other organisms. Our expected concentration values are similar to values that Mallin *et al.* (2004) reported as effective in increasing productivity of phytoplankton in tidal creeks.

Therefore, we may expect that our nitrate deposition would indeed provide at least a small fertilizer effect.

Overall, we can conclude that in general astrophysical ionizing radiation events are not likely to contribute large enough nitrate concentrations to have a significant additional negative impact on amphibian populations in combination with already considered UVB enhancement, and may actually provide some small benefit by acting as fertilizer. Further studies on the negative effects of these events can therefore concentrate on better predicting the impact of increased UVB alone.

5. **Acknowledgements**

The authors wish to thank the following individuals for helpful personal communications: Adrian Melott, Andrew Blaustein, Stephen Angel and David Henderson. BCT acknowledges support from a Washburn University Small Research Grant.

**References**

Bancroft, B., N.J. Baker, and A.R. Blaustein (2007) Effects of UVB radiation on marine and freshwater organisms: a synthesis through meta-analysis, *Ecology Lett.* 10, 332-345.

Belden, L.K. and Blaustein, A.R. (2002) Population differences in sensitivity to UV-B radiation for larval long-toed salamanders, *Ecology*, 83, 1586-1590.

Blaustein, A.R., Hoffman, P.D., Hokit, D.G., Kiesecker, J.M., Walls, S.C., and Hays, J.B. (1994) UV-repair and resistance to solar UV-B in amphibian eggs: A link to population declines? *Proc. Nat. Acad. of Sci. USA*, 91, 1791-1795.

Blaustein, A.R. and Wake, D.B. (1995) The puzzle of declining amphibian populations, *Scientific American* 272, 52-57.

Blaustein, A.R. and Kiesecker, J.M. (2002) Complexity in conservation: lessons from the global decline of amphibian populations, *Ecology Letters* 5, 597, doi:10.1046/j.1461-0248.2002.00352.x

Boone, M.D., Semlitsch, R.D., Little, E.E., and Doyle, M.C. (2007) Multiple stressors in amphibian communities: Effects of chemical contamination, bullfrogs and fish, *Ecol. App.* 17, 291-301.

Ejzak, L.M., Melott, A.L., Medvedev, M.V., and Thomas, B.C. (2007) Terrestrial Consequences of Spectral and Temporal Variability in Ionizing Photon Events, *Astrophys. J.* 654, 373-384.

Hatch, A.C. and Blaustein, A.R. (2000) Combined Effects of UV-B, Nitrate and Low pH Reduce Survival and Activity Level of Larval Cascades Frogs (*Rana cascadae*), *Arch. Environ. Contam. Toxicol.* 39, 494-499.

Hatch, A.C. and Blaustein, A.R. (2003) Combined Effects of UV-B Radiation and Nitrate Fertilizer on Larval Amphibians, *Ecological Applications* 13, 1083-1093.

Johansson, M., Rasanen, K. and Merila, J. (2001) Comparison of nitrate toleration between different populations of the common frog, *Rana temporaria*, *Aquatic Toxicology* 54, 1-14.

Mallin, M.A., *et al*. (2004) Nutrient limitation and algal blooms in urbanizing tidal creeks, *J. Exp. Marine Biology and Ecology*, 298, 211-231, doi:10.1016/S0022-0981(03)00360-5

Melott, A.L. *et al*. (2005) Climatic and Biogeochemical Effects of a Galactic Gamma-Ray Burst, *Geophys. Res. Lett*. 32, L14808 doi:10.1029/2005GL023073

Shi, Y., Hu, H. and Cong, W. (2005) Positive effects of continuous low nitrate levels on growth and photosynthesis of Alexandrium tamarense (Gonyaulacales, Dinophyceae), *Phycological Res.*, 53, 43–48. doi:10.1111/j.1440-183.2005.00371.x

Smith, V.H., Tilman, G.D. and Nekola, J.C. (1999) Eutrophication: impacts of excess nutrient inputs on freshwater, marine, and terrestrial ecosystems, *Environmental Pollution*, 100, 179-196, doi:10.1016/S0269-7491(99)00091-3

Setlow, R.B. (1974) The wavelengths in sunlight effective in producing skin cancer: A theoretical analysis, *Proc. Nat. Acad. Sci. USA,* 71, 3363-3366.

Thomas, B.C. *et al.* (2005) Gamma-ray bursts and the Earth: Exploration of Atmospheric, Biological, Climatic, and Biogeochemical Effects, *Astrophys. J.* 634, 509-533.

Thomas, B.C. & Melott, A.L. (2006) Gamma-ray bursts and terrestrial planetary atmospheres, *New Journal of Physics* 8, 120   doi:10.1088/1367-2630/8/7/120

Thomas, B.C., Jackman, C.H., and Melott, A.L. (2007) Modeling atmospheric effects of the September 1859 Solar Flare, *Geophys. Res. Lett.* 34, L06810, doi:10.1029/2006GL029174

Thomas, B.C., Melott, A.L., Fields, B.D., and Anthony-Twarog, B.J. (2008) Superluminous Supernovae: No Threat from eta Carinae, *Astrobiology* 8, 9-16.

**Tables**

Table 1: Conversion factor values computed for each site over 10 years. "NA" indicates data is not available for that site and year.

| Sites | 1994 | 1995 | 1996 | 1997 | 1998 | 1999 | 2000 | 2001 | 2002 | 2003 | 2004 | Average | Stdev |
|---|---|---|---|---|---|---|---|---|---|---|---|---|---|
| Loudonville | 0.12 | 0.13 | 0.12 | 0.13 | 0.13 | 0.13 | 0.09 | 0.14 | 0.11 | 0.10 | 0.09 | **0.12** | **0.02** |
| Westfield | 0.11 | 0.12 | 0.08 | 0.10 | 0.13 | 0.11 | 0.09 | 0.11 | 0.08 | 0.08 | 0.08 | **0.10** | **0.02** |
| Elmira | 0.12 | 0.14 | NA | NA | 0.13 | 0.16 | 0.13 | 0.15 | 0.11 | 0.10 | 0.10 | **0.13** | **0.02** |
| Buffalo | 0.13 | 0.14 | 0.09 | 0.11 | 0.12 | 0.12 | 0.08 | 0.12 | 0.09 | 0.11 | 0.11 | **0.11** | **0.02** |
| Whiteface Mt. | NA | 0.11 | 0.09 | 0.10 | 0.09 | 0.11 | 0.08 | 0.09 | 0.08 | 0.09 | 0.09 | **0.09** | **0.01** |
| Piseco Lake | 0.10 | 0.10 | 0.07 | 0.10 | 0.11 | 0.11 | 0.07 | 0.10 | 0.08 | 0.07 | 0.08 | **0.09** | **0.01** |
| Nicks Lake Campground | 0.09 | 0.10 | 0.08 | 0.10 | 0.09 | 0.10 | 0.08 | 0.09 | 0.08 | 0.08 | 0.08 | **0.09** | **0.01** |
| Camp Georgetown | 0.11 | 0.13 | 0.09 | 0.13 | 0.00 | 0.12 | 0.08 | 0.10 | 0.07 | NA | 0.08 | **0.09** | **0.04** |
| Rochester | 0.15 | 0.15 | 0.10 | 0.13 | 0.11 | 0.17 | 0.13 | 0.11 | 0.13 | 0.13 | NA | **0.13** | **0.02** |
| Eisenhower Park | 0.13 | 0.11 | 0.08 | 0.10 | 0.11 | 0.11 | 0.09 | 0.12 | 0.08 | 0.08 | 0.09 | **0.10** | **0.02** |
| Niagara Falls | 0.15 | 0.15 | NA | NA | NA | 0.14 | 0.12 | 0.16 | 0.13 | 0.15 | 0.13 | **0.14** | **0.01** |
| East Syracuse | 0.13 | 0.15 | 0.11 | 0.17 | 0.13 | 0.15 | 0.10 | 0.11 | 0.09 | 0.09 | 0.10 | **0.12** | **0.03** |
| Altmar | 0.10 | 0.10 | 0.08 | 0.10 | 0.10 | 0.10 | 0.09 | 0.11 | 0.09 | NA | 0.09 | **0.10** | **0.01** |
| Mt. Ninham | 0.09 | 0.10 | 0.07 | 0.10 | 0.10 | 0.11 | 0.09 | 0.11 | 0.08 | 0.06 | 0.08 | **0.09** | **0.02** |
| Wanakena Ranger School | 0.11 | 0.10 | 0.08 | 0.09 | 0.09 | 0.10 | 0.08 | 0.09 | 0.10 | 0.09 | 0.10 | **0.09** | **0.01** |
| Belleayre Mt. | 0.10 | 0.11 | 0.07 | NA | 0.09 | 0.09 | 0.08 | 0.13 | 0.09 | 0.04 | 0.08 | **0.09** | **0.02** |
| White Plains | 0.08 | 0.10 | 0.07 | 0.09 | 0.10 | 0.08 | 0.09 | 0.10 | 0.08 | 0.07 | 0.07 | **0.08** | **0.01** |
| New York Botanical Gardens | NA | 0.10 | 0.08 | 0.09 | 0.09 | 0.10 | 0.10 | 0.10 | 0.08 | 0.07 | 0.06 | **0.09** | **0.02** |